# Optomechanics with one-dimensional gallium phosphide photonic crystal cavities

Katharina Schneider, Yannick Baumgartner, Simon Hönl, Pol Welter, Herwig Hahn, Dalziel J. Wilson, Lukas Czornomaz, Paul Seidler*

*pfs@zurich.ibm.com

IBM Research – Zurich, Säumerstrasse 4, CH-8803 Rüschlikon, Switzerland

**Gallium phosphide offers an attractive combination of a high refractive index ($n > 3$ for vacuum wavelengths up to 4 μm) and a wide electronic bandgap (2.26 eV), enabling optical cavities with small mode volumes and low two-photon absorption at telecommunication wavelengths. Heating due to strongly confined light fields is therefore greatly reduced. Here, we investigate the benefits of these properties for cavity optomechanics. Utilizing a recently developed fabrication scheme based on direct wafer bonding, we realize integrated one-dimensional photonic crystal cavities made of gallium phosphide with optical quality factors as high as $1.1 \times 10^5$. We optimize their design to couple the optical eigenmode at ~200 THz via radiation pressure to a co-localized mechanical mode with a frequency of 3 GHz, yielding sideband-resolved devices. The high vacuum optomechanical coupling rate ($g_0 = 2\pi \times 400$ kHz) permits amplification of the mechanical mode into the so-called mechanical lasing regime with input power as low as ~20 μW. The observation of mechanical lasing implies a multiphoton cooperativity of $C > 1$, an important threshold for the realization of quantum state transfer protocols. Because of the reduced thermo-optic resonance shift, optomechanically induced transparency can be detected at room temperature in addition to the normally observed optomechanically induced absorption.**

## 1. Introduction

Optical microcavities possessing mechanical resonances with frequencies of several gigahertz hold the potential for creating electro-opto-mechanical devices with applications as diverse as the transduction of microwave qubits into optical qubits [1,2], the detection of radio signals in astronomy [3], and low-power, high-speed data communication [4]. A platform for such gigahertz-frequency systems is provided by one- and two-dimensional photonic crystal cavities (PhCCs) [5,6], which have previously been made of materials such as silicon [5], silicon nitride [7], gallium arsenide [8], aluminum nitride [2] and diamond [9]. As photonic crystals rely on index contrast, a high refractive index $n$ is a beneficial material property. To reduce optical losses and prevent heating of the device, a wide electronic bandgap $E_g$ is also desirable because, if sufficiently large, it can reduce two-photon absorption at typically used near-infrared wavelengths. These two requirements are inherently conflicting: a large refractive index implies high polarizability at optical frequencies [10], a property usually associated with relatively weakly bound electrons, whereas a wide bandgap results from strong electronic bonds between atoms [11]. Of the aforementioned materials, only AlN ($E_g = 6.2$ eV [12], $n = 2.03$ [13]), diamond ($E_g = 5.4$ eV [11], $n = 2.39$ [14]) and Si$_3$N$_4$ ($E_g \approx 5$ eV [15], $n = 2.00$ [16]) have a large enough bandgap to prevent two-photon absorption at a vacuum wavelength of $\lambda_\mathrm{vac} = 1550$ nm, but at the cost of a refractive index well below 3.

As an alternative for optomechanics, we introduce photonic crystal cavities made of gallium phosphide ($E_g = 2.26$ eV [11], $n = 3.05$ [17]) [18]. The index of refraction of GaP is the highest among visibly transparent binary semiconductors. In addition, its non-centrosymmetric crystal structure offers the possibility of piezoelectric actuation. Optomechanical coupling has been previously observed in microdisks made of GaP [19]. PhCCs have the advantage of possessing smaller mode volume, but the few examples of two-dimensional PhCCs that have been fabricated out of GaP [20] have not been investigated for their optomechanical behavior. Profiting from a new fabrication process for chip-level integration of GaP photonic devices based on direct wafer bonding onto low-refractive-index substrates [21], we report the first investigation of optomechanics with a PhCC made of GaP.

Optomechanical coupling occurs between a mechanical mode with resonance frequency $\Omega_\mathrm{m}$ spatially overlapping an optical eigenmode with resonance frequency $\omega_0$ [22]. Mechanical deformation of the device (described as a generalized displacement $x$), leads to a change of the optical resonance frequency. For small displacements, this change can be approximated with a linear dependence characterized by the vacuum optomechanical coupling rate $g_0 = \frac{\partial \omega_0}{\partial x} \cdot x_\mathrm{zpf}$, where $x_\mathrm{zpf} = \sqrt{(\hbar/2m_\mathrm{eff}\Omega_\mathrm{m})}$ is the zero-point motion, $m_\mathrm{eff}$ is the effective mass, and $\hbar$ the reduced Planck constant. Efficient transduction of a photon to a quantum of mechanical vibration and vice versa requires a high cooperativity $C = n_\mathrm{cav}\frac{4g_0^2}{\kappa\Gamma}$, where $n_\mathrm{cav}$ is the number of intracavity photons. The decay rates of the optical and mechanical resonances, $\kappa$ and $\Gamma$, respectively, are therefore quantities to be minimized. Furthermore, cooling or amplification of the mechanical mode is most efficient in the so-called resolved-sideband regime, where $\Omega_m \geq \kappa$.

Our one-dimensional GaP PhCCs have optical quality factors as high as $Q_o = \omega_o/\kappa = 1.1 \times 10^5$ ($\kappa$ as low as $2\pi \times 1.8$ GHz). The devices are optimized to couple the optical mode to a mechanical mode with a resonance frequency of $\Omega_m = 2\pi \times 2.9$ GHz, putting the devices in the resolved-sideband regime, and an optomechanical coupling rate of $g_0 = 2\pi \times 400$ kHz is achieved. The device can be driven into "mechanical lasing" [23] at room temperature, which implies a multiphoton cooperativity $C > 1$, with input power as low as 20 μW. In addition to the usual optomechanically induced absorption (OMIA), we observe optomechanically induced transparency (OMIT) [24,25] at room temperature. This is only possible because the thermo-optic shift of the optical resonance is sufficiently small to allow red-detuning of the control laser with respect to the cavity resonance on the order of $\Omega_m$. To date, OMIT at room temperature has only ever been shown for diamond [9] and Si$_3$N$_4$ devices [7].

## 2. Methods

### A. Design

The PhCC consists of a freestanding GaP beam patterned with elliptical holes. The hole radii increase quadratically with hole number from the center of the device outward, followed by a set of identical holes, as illustrated in Fig 1(a). In this way, gradual mode matching is achieved between the center of the cavity and the Bragg-mirror outer portion of the structure. To attain a design which supports co-localized optical and mechanical modes with a high optomechanical coupling rate, numerical simulations with the finite-element solver COMSOL Multiphysics [26] were performed. The COBLYA algorithm [27], which is part of the NLOPT package [28], was employed varying those dimensions of the PhCC indicated in Fig. 1(b) for a GaP layer thickness of 300 nm and maximizing the fitness, defined as $F = g_0 \cdot Q_o$ (with $Q_o$ capped at $1.5 \times 10^6$ to avoid unrealizably high values) [29]. Both moving boundary [30] and photoelastic [31] contributions to the optomechanical coupling rate were considered, where the long axis of the PhCC is oriented in the [110] crystal direction (see Supplement 1 for details). The resulting optimized structure (ellipse radii $r_a$ and $r_b$ and unit cell size $a$ shown in Fig. 1(a) and $w = 542$ nm) has an optical quality factor $Q_o > 1.4 \times 10^6$ and a mode volume of 0.0971 $(\lambda_{vac}/n)^3$ at $\lambda_{vac} = 1493$ nm. The mechanical breathing mode, for which the displacement profile is shown in Fig. 1(c), is calculated to have a frequency of $\Omega_m = 2\pi \times 2.84$ GHz, an effective mass of $m_{eff} = 8.6 \times 10^{-13}$ g, and an optomechanical coupling rate of $g_0 = 2\pi \times 760$ kHz. The photoelastic contribution ($g_{pe} = 2\pi \times 725$ kHz) dominates the moving boundary contribution of the same sign ($g_{mb} = 2\pi \times 35$ kHz) because the motion-induced strain profile overlaps well with the regions of high electric field (compare Fig. 1(d) and (e)).

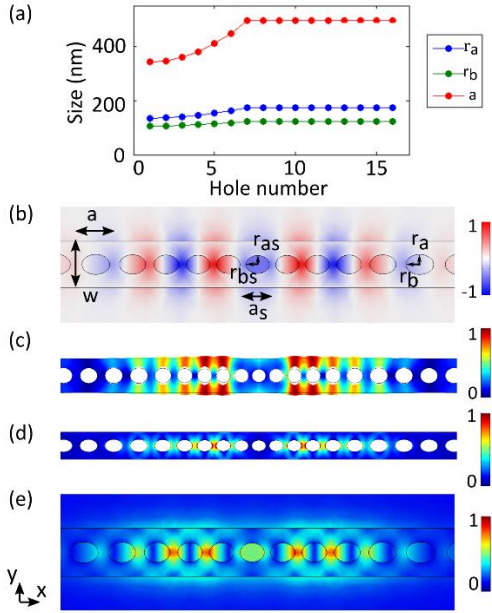

Fig. 1. Optimized device design obtained from finite-element simulations. (a) Dimensions of elliptical holes and unit cell. (b) Electric field y-component. The overlaid outline of the device structure is labelled with the dimensions optimized in the finite-element simulations. (c) Displacement profile of the mechanical breathing mode at 2.84 GHz. (d) Principal strain for the breathing mode. (e) Magnitude of the electric field. Figures (b) – (e) are all for a plane passing through the middle of the PhCC and the color scales are normalized.

### B. Fabrication

To fabricate the PhCCs, we build on our previously published method for realizing integrated GaP-on-insulator (GaPoI) devices [21], with some modifications. Details are given in Supplement 1. Briefly, an epitaxial GaP/Al$_x$Ga$_{1-x}$P/GaP heterostructure is grown by metal-organic chemical vapor deposition on a [100]-oriented GaP wafer, which is subsequently bonded to an oxidized silicon wafer. The original GaP growth wafer is first thinned by lapping and then dry etched with SiCl$_4$ and SF$_6$ in a process that selectively removes GaP in the presence of Al$_x$Ga$_{1-x}$P [32]. The Al$_x$Ga$_{1-x}$P etch-stop layer is removed selectively with concentrated HCl, yielding the desired GaPoI substrate. The PhCCs, which are oriented along the [110] crystal direction, as well as the waveguides and grating couplers to which they are connected, are patterned in the GaP device layer via e-beam lithography and a chlorine-based dry etching technique that has been optimized for geometric and dimensional accuracy. The PhCCs are released to form freestanding structures in a final step by wet etching with buffered hydrofluoric acid to remove ~1.1 µm of the underlying SiO$_2$, during which the connecting waveguides and grating couplers are protected by photoresist.

A scanning electron microscope (SEM) image of a PhCC thus fabricated is shown in Fig. 2. The devices feature smooth, nearly vertical sidewalls, as evident from a cross-sectional image prepared by focused-ion-beam milling (Fig. 2(c)). Faithful reproduction of the simulated design is essential, as the optical quality factor drops by 70% if the hole radii are too large by 10% (see Supplement 1).

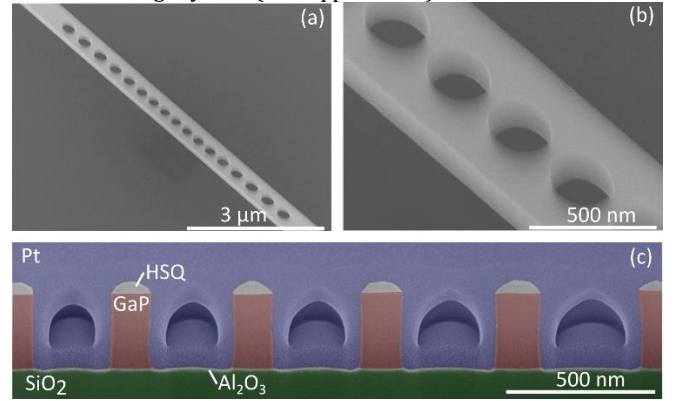

Fig. 2. SEM images of a one-dimensional GaP PhCC. (a) Freestanding device with width and height of 542 nm and 300 nm, respectively. (b) Magnification of the central part of the PhCC showing smooth and straight sidewalls. (c) Device cross-section before release, prepared by focused-ion-beam milling. False coloring indicates different materials.

### C. Characterization

The apparatus used to characterize the optical and mechanical properties of our PhCCs is shown in Fig 3. For ease of interrogation, the PhCCs are connected to waveguides terminated with grating couplers designed with a nominal coupling angle 10° from vertical [21]. Continuous-wave infrared light from a tunable external cavity laser (Photonetics Tunics-Plus) is directed through a cleaved single-mode optical fiber into the input grating coupler. Fiber polarization controllers (FPC) align the polarization of the light to the TE design orientation of the device. Light transmitted through the device is gathered with a second single-mode fiber positioned over the output grating coupler. A portion of the collected light is split off with a fiber beam splitter (FBS) to a power meter (EXFO IQ 1600) to monitor the power level. For measurements of the thermomechanical radio frequency spectrum, the transmitted light is amplified by an erbium-doped fiber amplifier (EDFA, JDS Uniphase MAP Amplifier) followed by a bandpass filter (BPF, JDS Uniphase TB3) and sent to an optical receiver consisting of a fast photodiode with a built-in transimpedance amplifier (JDS Uniphase MAP Receiver RX10). The transduced photocurrent from the photodiode is evaluated with an electrical spectrum analyzer (HP8563A). For measurements of optomechanically induced absorption and transparency, the laser output is phase modulated with an electro-optic modulator (EOM, Thorlabs 10 GHz LN65S-FC) before entering the device. A frequency generator (HP 8341A Synthesized Sweeper 0.01 – 20 GHz) delivers the radio frequency input signal for the

EOM, and a vector network analyzer (HP8510B) monitors the frequency dependent response of the device.

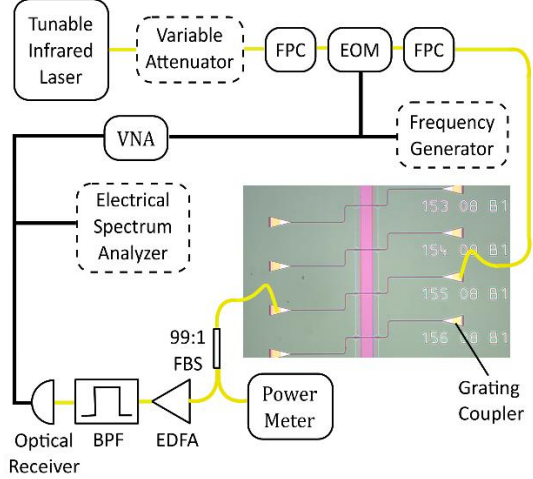

Fig. 3. Schematic of the measurement apparatus. Dashed lines indicate optional equipment used for certain measurements. In the optical microscope image, the pink stripe is where ~1.1 µm of the underlying SiO$_2$ layer has been removed by wet etching to release the PhCCs.

## 3. Results

### A. Optical transmission

By fitting a Lorentzian function to the transmission spectrum of a PhCC, we obtain the optical quality factor $Q_o = \omega_o/\kappa$, where $\omega_o$ is the optical resonance frequency and $\kappa$ the Lorentzian linewidth. The highest quality factor measured, $Q_o = 111000 \pm 2000$, is for a device with 12 holes on each side (Fig. 4 (a)) and is close to the intrinsic quality factor $Q_i$, as is apparent from Fig. 4 (b). With increasing number of holes, the coupling rate to the attached waveguides decreases, the external quality factor $Q_e$ becomes larger than $Q_i$, and the measured quality factor $Q_o = \left(\frac{1}{Q_e} + \frac{1}{Q_i}\right)^{-1}$ approaches $Q_i$.

The typical power dependence of the transmission spectrum of a PhCC is shown in Fig. 4(c), in this case for a device with 11 holes on each side. With increasing intracavity photon number, the optical resonance shifts to lower frequency (Fig. 4(e)) due to both the thermo-optic and Kerr effect, where the former most likely dominates [33]. For many experiments in optomechanics, it is desirable to reduce this shift as much as possible, as it can severely constrain the ability to independently control laser detuning with respect to the cavity resonance for high intracavity energy levels. Heating due to absorption also limits the lowest attainable temperature in cryogenic experiments. A plot of the quality factors $Q_o$ (Fig. 4(d)) obtained from the power-dependent spectra in Fig. 4(c) reveals an increase with input power, implying a decrease in the internal optical losses. Two general regimes are observed, one at the lowest powers where $Q_o$ changes rapidly with increasing power, and a second in which $Q_o$ increases gradually and eventually plateaus. A similar increase in quality factor with input power has been reported previously for GaP microdisks and tentatively attributed to the presence of saturable absorbers [19]. For two-photon absorption, as occurs with silicon [34], the opposite, namely a decrease of $Q_o$ with increasing power is expected.

### B. Thermo-optic behavior

To obtain a more quantitative measure of the thermo-optic effect, we analyze the dependence of the optical resonance frequency on the power dissipated in the cavity $P_d = \kappa_i U_{\text{cav}}$ (Fig. 4(e)) [35,36], the derivative of which we define as the thermal susceptibility $\chi_{\text{th}} = \frac{d\omega_o}{dP_d}$. Here, $\kappa_i$ is the intrinsic loss rate of the cavity and $U_{\text{cav}}$ is the intracavity energy. We are interested in the fraction $\zeta$ of the intrinsic loss rate that comes from absorption and leads to an increase in temperature of the device, namely $\kappa_{\text{abs}} = \zeta \kappa_i$. Assuming the frequency shift $\delta\omega_o$ is entirely due to thermo-optic effects, we can write $\delta\omega_o = \beta(\omega_o)\Delta T$, where $\beta(\omega_o)$ is a frequency-dependent coefficient incorporating both the thermo-refractive effect and thermal expansion, and $\Delta T$ is the local temperature increase. The power absorbed, $P_{\text{abs}} = \kappa_{\text{abs}} U_{\text{cav}}$, is also the heat flow through the device, and thus $\Delta T = R_{\text{th}} P_{\text{abs}}$, where $R_{\text{th}}$ is the thermal resistance of the structure. Combining these equations gives $\kappa_{\text{abs}} = \kappa_i \chi_{\text{th}}/\beta(\omega_o) R_{\text{th}}$.

Taking a linear fit of the higher power portion of the plot of frequency shift versus power dissipated in the cavity (see Fig. 4(e)), we estimate $\chi_{\text{th}} = -2\pi \times 0.76$ THz/mW for this particular device, the magnitude of which we consider an upper limit given that the quality factor has not yet plateaued at these power levels. We determine $\beta(\omega_o) = -0.00683$ THz/K directly by measuring the optical resonance frequency as a function of sample temperature (see Supplement 1). In contrast, $R_{\text{th}}$ is not easily determined experimentally, but from finite element simulations we estimate a value of $1.93 \times 10^3$ K/mW (see Supplement 1). We then have $\kappa_{\text{abs}} \approx 2\pi \times 0.12$ GHz. For comparable devices made of silicon [37], we find $\kappa_{\text{abs}} \approx 2\pi \times 0.60$ GHz and $\kappa_{\text{abs}} \approx 2\pi \times 0.84$ GHz. We conclude that $\kappa_{\text{abs}}$ is at least a factor of 5 smaller in the GaP PhCCs for power levels at which the optical quality factor is no longer changing. The thermo-optic frequency shift $\partial\omega = \beta(\omega_o) R_{\text{th}} \kappa_{\text{abs}} U_{\text{cav}}$ expected for a given intracavity energy $U_{\text{cav}}$ should also be at least a factor of 5 smaller in the GaP PhCC, as both $R_{\text{th}}$ and $\beta(\omega_o)$ are nearly the same for the GaP and Si devices considered here. (See Supplement 1 for details.)

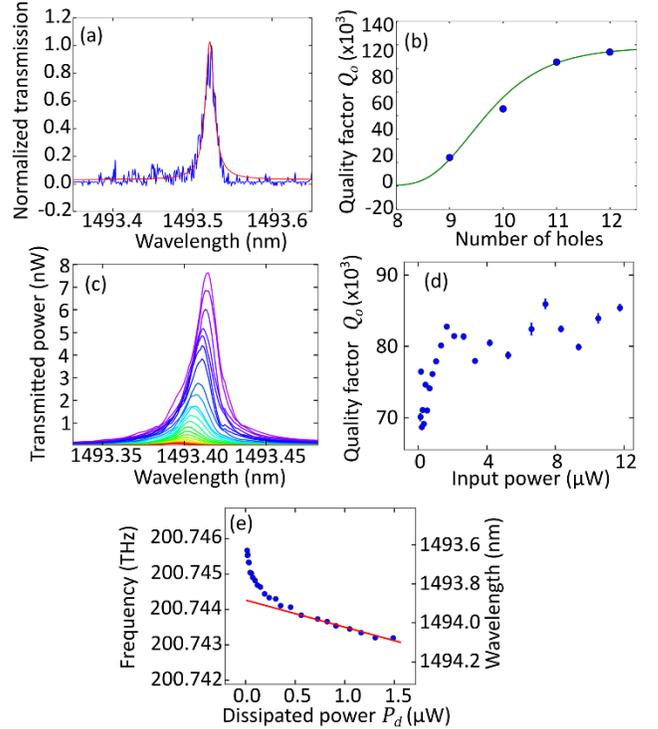

Fig. 4. Analysis of optical transmission measurements. (a) Spectrum of a PhCC with the highest measured optical quality factor $Q_o = (1.11 \pm 0.02) \times 10^5$. (b) Dependence of $Q_o$ on the number of holes on each side of the PhCC. The green curve is a guide to the eye. (c) Transmission spectra of a device with 11 holes on each side for input powers ranging from 0.079 µW to 7.7 µW. (d) Increase of the optical quality factor with input power. (e) Dependence of the resonance frequency on dissipated power $P_d$. The red line is a linear fit to the data for dissipated power ≥ 0.55 µW.

The above conclusions hold when the circulating power in the cavity is sufficient for the quality factor to have reached a steady value. For the GaP device examined here, this corresponds to less than a thousand intracavity photons, well below the level used in many optomechanics experiments. We also note that the same general behavior as shown in Fig. 4(c) – (e) was observed for several devices. Measurement in a nitrogen atmosphere did not have an effect.

### C. Thermomechanical fluctuations

Thermal motion of the GaP PhCC leads to phase modulation of the transmitted light. The resulting thermomechanical radio frequency spectrum is obtained by monitoring the power fluctuations of a detuned probe field using a fast photoreceiver and an electrical spectrum analyzer. An example is shown in Fig. 5 for a device with ten holes on each side. Each mechanical mode of the PhCC produces a Lorentzian-shaped peak, the amplitude of which is proportional to the vacuum optomechanical coupling rate $g_0$ for that mode. The breathing mode for which the device design was optimized is found at 2.902 GHz. Another strongly coupled mode is observed at 130 MHz and is assigned on the basis of the numerical simulations to an accordion mode, in which the nanobeam stretches longitudinally such that all holes are displaced symmetrically with respect to the center of the device (see the color-coded images in Fig. 5).

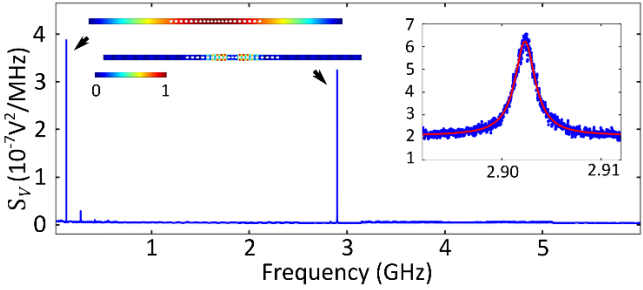

Fig. 5. Thermomechanical radio frequency spectrum of a device with ten holes on each side. The two dominant modes are the accordion mode at 130 MHz and the breathing mode at 2.902 GHz. Simulated displacement profiles with a normalized color scale are shown for each mode. An expanded view of the breathing mode resonance is given in the inset.

### D. Vacuum optomechanical coupling rate

To quantify the vacuum optomechanical coupling rate $g_0$, two independent measurements are performed. In the first, we compare the power spectral density resulting from the thermomechanical cavity frequency fluctuation with that produced by a calibration tone [38]. We generate the calibration tone by phase-modulating the laser light entering the cavity at a frequency near that of the mechanical resonance and with a known modulation depth. Comparing the area under the thermomechanical peak to that under the calibration peak gives $g_0 = 2\pi \times (370 \pm 40)$ kHz. Details are provided in Supplement 1.

The second method exploits optomechanically induced absorption (OMIA) [24,25]. In this case, the laser light is weakly phase modulated prior to entering the cavity, resulting in a strong control field at the laser frequency $\omega_c$ with weak sidebands that serve as probe fields at frequency $\omega_p$. Constructive interference of the light mechanically modulated in the cavity with the probe field at frequency-matched detuning results in the enhanced transmission window known as OMIA. To maximize the OMIA peak, the control field is blue-detuned from the cavity resonance by the mechanical resonance frequency, i.e. $\Delta_{oc} = \omega_o - \omega_c = \Omega_m$. The probe-control detuning $\Delta_{pc} = \omega_p - \omega_c$ is varied by changing the phase modulator drive frequency, which is provided by the output port of a vector network analyzer (VNA). The resulting transmission is evaluated by routing the output of the photoreceiver to the response port of the VNA and performing an $S_{21}$ measurement. A coarse sweep of the modulation frequency over 10 GHz maps out the optical resonance of the PhCC, as shown in Fig. 6(a) for the same device as used for the calibration-tone measurement above. A finer sweep around the maximum of the $|S_{21}|$ signal (Fig. 6(b)) reveals the OMIA peak at $\Omega_m = 2\pi \times 2.902$ GHz. From a fit of the expected transmission profile (see Supplement 1 for details), we infer $G = \sqrt{n_{\text{cav}}} \cdot g_0$ [39,40]. With knowledge of the optical power entering and leaving the grating couplers, the average grating-coupler losses, and the laser-cavity detuning $\Delta_{oc}$, the intracavity photon number $n_{\text{cav}}$ can be estimated [41]. For the measurement shown in Fig. 6(a) and (b), $n_{\text{cav}} = 2100 \pm 700$. Averaging over several measurements with different detunings $\Delta_{oc}$ gives $g_0 = 2\pi \times (680 \pm 320)$ kHz, in agreement with the result obtained in the calibration-tone experiment, albeit with greater uncertainty.

We believe the calibration-tone method provides a more reliable value for $g_0$ than the OMIA technique because it does not require knowledge of the intracavity photon number, which is difficult to precisely estimate in our experiment. The measured coupling rate differs from the simulated value by approximately a factor of two. This may be due to fabrication inaccuracies or an incorrect estimate of the photoelastic coefficients. Both of these potential causes of error are treated in Supplement 1.

### D. Optomechanically induced transparency

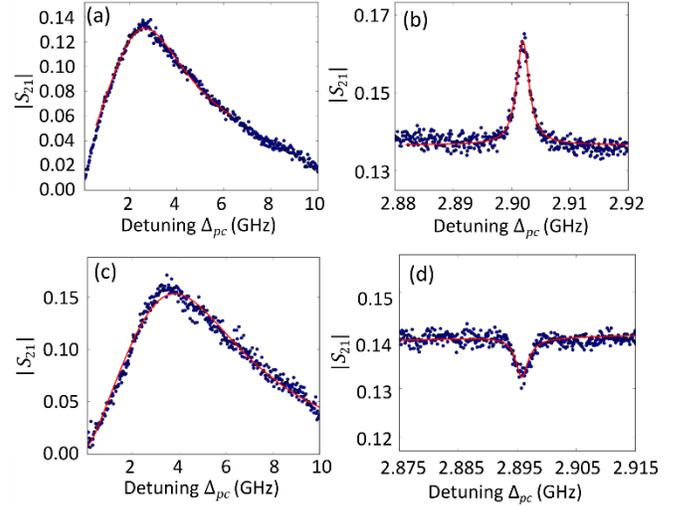

Fig. 6. Measured VNA traces (blue dots) with fits to model (red curves). (a) and (b) OMIA measurement with the laser control frequency blue-detuned from the optical resonance. (c) and (d) OMIT measurement with the laser control frequency red-detuned from the optical resonance. The OMIA peak and OMIT dip in (b) and (d), respectively, are not evident in the broadband spectra (a) and (c) because of the limited number of data points.

Optomechanically induced transparency (OMIT) is analogous to OMIA, only the laser is red-detuned with respect to the cavity, i.e. $\Delta_{oc} = -\Omega_m$, such that the probe and control fields interfere destructively with the mechanically modulated laser light [24,25]. This leads to decreased transmission at $\Delta_{pc} = \Omega_m$ (Fig. 6(c) and (d)). In order to carry out this measurement, the input power must be set sufficiently low that the thermo-optic shift is small and the red-detuning near $\Delta_{oc} = -\Omega_m$ is stable. This is usually not possible at room temperature with silicon PhCCs because of the pronounced thermo-optic bistability. OMIT has however been observed at room temperature for PhCCs made of diamond [9] and Si$_3$N$_4$ [7], which like GaP have a wide electronic bandgap and therefore weak two-photon absorption. Diamond has the additional advantage of high thermal conductivity, further mitigating heating [42]. We observed OMIT in a device with a degraded optical quality factor (possibly due to photo-degradation) that was no longer in the resolved-sideband regime (Fig. 6(c) and (d)). Consequently, the measured OMIT dip is shallow compared to the magnitude of the OMIA peak in Fig. 6(b) because of the need to use lower laser input power and operate at $|\Delta_{oc}| > \Omega_m$ in order to ensure stable detuning.

### E. Dynamical backaction

If the PhCC is driven off resonance, radiation pressure gives rise to stiffening (or softening) and damping (or amplification) of its mechanical modes [43]. We measured these effects only for blue-detuning of the laser, as only on the blue side of the optical resonance could we vary the detuning continuously without contending with thermo-optic bistability. The expected frequency shift $\Delta\Omega_m$ and change in linewidth $\Gamma_{opt}$ induced by the light field are [43]

$$\delta\Omega_m = g_0^2 \bar{n}_{cav} \left( \frac{\Delta_{oc} - \Omega_m}{\kappa^2/4 + (\Delta_{oc} - \Omega_m)^2} + \frac{\Delta_{oc} + \Omega_m}{\kappa^2/4 + (\Delta_{oc} + \Omega_m)^2} \right) \quad (1)$$

$$\Gamma_{opt} = g_0^2 \bar{n}_{cav} \left( \frac{\kappa}{\kappa^2/4 + (\Delta_{oc} + \Omega_m)^2} - \frac{\kappa}{\kappa^2/4 + (\Delta_{oc} - \Omega_m)^2} \right). \quad (2)$$

We determined the resulting effective mechanical resonance frequency $\Omega_{eff} = \Omega_m + \delta\Omega_{opt}$ and effective mechanical linewidth $\Gamma_{eff} = \Gamma_m + \Gamma_{opt}$ ($\Gamma_m$ is the intrinsic mechanical linewidth) from the thermomechanical radio frequency spectrum while varying the laser-cavity detuning $\Delta_{oc}$. Because of the thermo-optic shift of the optical resonance, it is not straightforward to determine $\Delta_{oc}$ for a given laser frequency $\omega_c$. We used OMIA spectra and the transmitted power as two independent measurements from which to infer the detuning (see Supplement 1 for further details). We calculated $\Delta\Omega_m$ and $\Gamma_{opt}$ from the average of the two values thus determined. The results, which demonstrate the presence of both the optical spring effect and parametric amplification, are displayed in Fig. 7(a) and (b). The red curves show the expected backaction according to Eq. (1) and (2) with $g_0 = 2\pi \times 400$ kHz taken from the calibration tone measurement (section C) and $\kappa = 2\pi \times 3.15$ GHz taken from the optical transmission profile. For this particular experiment, the intracavity photon number was estimated from the measured power leaving the cavity to be $n_{cav} = 10000 \pm 3300$ on resonance and assumed to have a Lorentzian distribution with detuning corresponding to the optical resonance. The model yields an intrinsic mechanical linewidth of $\Gamma_m = 2\pi \times 3.12$ MHz and an intrinsic mechanical frequency of $\Omega_m = 2\pi \times 2.9023$ GHz. The corresponding mechanical quality factor $Q_m \approx 930$ is similar to that for one-dimensional PhCCs made of other crystalline materials (e.g. Si [5,44], diamond [9], and GaAs [8]) operated at room temperature and atmospheric pressure.

If the detuning $\Delta_{oc}$ is kept constant and instead the intracavity photon number is varied, the mechanical frequency and linewidth change linearly with $n_{cav}$, which is shown in Fig. 7(c) and (d). Constant detuning was obtained by adjusting $\Delta_{oc}$ to maximize the height of the OMIA peak. As this implies a detuning of $\Delta_{oc} = \Omega_m$, we expect the optical spring effect to be negligible [22]. The intercepts of the linear fits to the mechanical linewidth and mechanical frequency give an intrinsic mechanical decay rate of $\Gamma_m = 2\pi \times (3.13 \pm 0.06)$ MHz and an intrinsic mechanical frequency $\Omega_m = 2\pi \times (2.90 \pm 0.03)$ GHz, respectively, consistent with the values obtained from the detuning-dependent measurements.

At even higher input power, the linewidth of the mechanical resonance narrows dramatically, signaling the onset of mechanical lasing (self-sustained oscillations, corresponding to $\Gamma_{opt} = -\Gamma_m$) [45–47]. The transition into this regime is shown in Fig. 7(e), where the detuning has been kept constant at $\Delta_{oc} \approx \Omega_m$ while increasing the power. The lowest observed input power threshold to achieve mechanical lasing was 20 µW for a device with ten holes on each side. Achieving mechanical lasing indicates that our system attains a photon-enhanced cooperativity $C = n_{cav} \frac{4 g_0^2}{\Gamma \kappa} > 1$ [22]. This is an important threshold: the transduction between light and mechanics is faster than the mechanical decoherence rate [43], which is a prerequisite for quantum state transfer protocols. Furthermore, the observation of self-sustained oscillations implies that, for equivalent red detuning, the temperature of the mechanical mode can be reduced by at least a factor of two [48].

Interestingly, the background of the thermomechanical spectrum drops in the transition to mechanical lasing. The cause of this behavior becomes apparent when the transmitted power is examined in the time domain with a high-speed oscilloscope (Tektronix CSA 8000). The amplitude changes from an almost constant value to a signal oscillating strongly at the mechanical frequency (see Fig. 7(f)). The non-sinusoidal shape is characteristic of limit-cycle oscillations and occurs when the magnitude of the cavity frequency shift is larger than $\kappa$ [47] [49]. The nonlinearity is also evident in the thermomechanical radio frequency spectrum (see Fig. 7(g)) as harmonics of the mechanical mode (limited to the fourth order by the photodiode bandwidth).

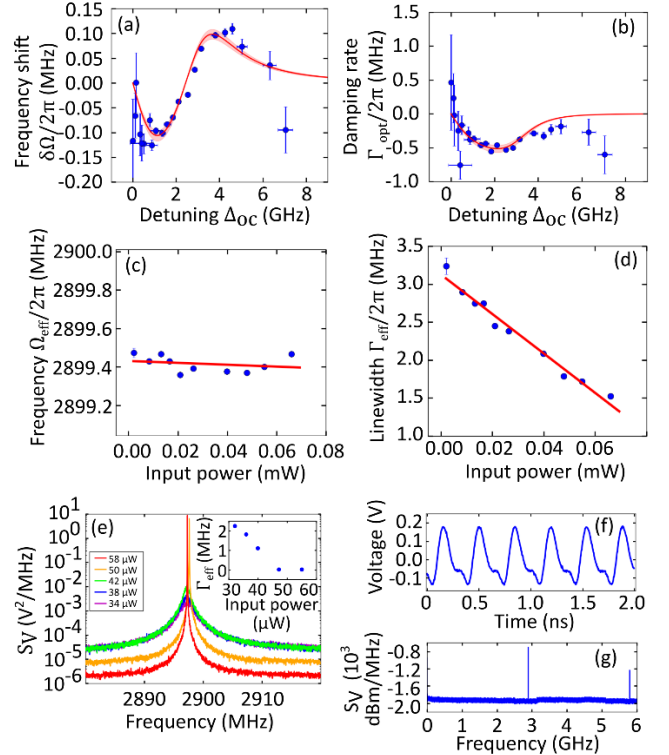

Fig. 7. Dynamical radiation pressure backaction. (a) and (b) Detuning-dependent optical stiffening and damping (blue dots are measured data; red curves are the prediction from the model, where the shaded area indicates uncertainty in $g_0$ of $\pm 2\pi \times 20$ kHz). (c) and (d) Power-dependent optical stiffening and damping (blue dots are measured data; red lines are the prediction from the model). (e) Dramatic linewidth narrowing of the mechanical resonance with increasing input power, where the inset shows the effective linewidth $\Gamma_{eff}$ obtained from a Lorentzian fit of the spectra. (f) Time-resolved transmission signal for 77 µW input power oscillating at a frequency of 2.91 GHz, i.e. the mechanical resonance frequency. The non-sinusoidal shape is due to the optomechanical backaction becoming nonlinear. The signal is effectively low-pass filtered at this frequency by the 10-GHz bandwidth of the photodiode. (g) Radio frequency spectrum showing the fundamental mechanical resonance and its first harmonic.

## 4. Conclusions

The work presented here is the first investigation of cavity optomechanics with PhCCs made of GaP. We have developed and fabricated a PhCC design that permits sideband-resolved coupling of an optical mode to a stationary mechanical mode with a frequency of 2.902 GHz. The fabricated devices exhibited optical quality factors as high as $Q_o = 1.1 \times 10^5$. The high vacuum optomechanical coupling rate of $g_0 = 2\pi \times 400$ kHz gives rise to phenomena such as OMIA, the optical spring effect, and parametric amplification, which we have characterized and compared to theoretical models. Efficient amplification of the mechanical mode by the optical pump manifested itself in the form of mechanical lasing with an input power threshold as low as ~20 μW. Mechanical lasing implies a multiphoton cooperativity $C > 1$, which is a prerequisite for quantum transduction and efficient cooling. We were also able to observe OMIT at room temperature because of weak thermo-optic bistability in comparison to that of comparable devices made of silicon. By measuring the power dependence of the optical mode frequency shift, we furthermore demonstrated that the wide electronic bandgap of GaP reduces two-photon absorption at typical telecommunication wavelengths. We thus confirm that GaP is advantageous for applications requiring high intracavity photon numbers while maintaining a stable optical resonance frequency.

A question that could not be fully resolved within the scope of this work is the apparent existence of saturable absorbers in the GaP, which leads to an initial increase in optical quality factor with increasing power. Also, the gradual degradation of the optical device properties over time is not yet fully understood. We have some indication from other work on ring resonators [33] that a protective coating, such as a few nanometers of $Al_2O_3$, may impede the deterioration, suggesting some sort of photooxidation as the underlying cause. Answers to both questions require more extensive study.

For future electro-opto-mechanical systems, the non-centrosymmetric crystal structure of GaP opens the door for potential piezoelectric actuation. Taken together with its low two-photon absorption and large index of refraction, which allow strong confinement of light with reduced heating at high optical powers, the material is a good candidate for microwave-to-optical transduction via an intermediate mechanical mode [2,50].

**Funding**. Marie Curie H2020-ETN OMT (722923) and FET Proactive HOT (732894)

**Acknowledgment**. We gratefully acknowledge Antonis Olziersky for the e-beam lithography, Daniele Caimi for the bonding of the wafers, Ute Drechsler for assistance with the ICP-RIE and Stefan Abel for the design of the grating couplers. This work was supported by the European Union's Horizon 2020 Program for Research and Innovation under grant agreement No. 722923 (Marie Curie H2020-ETN OMT) and No. 732894 (FET Proactive HOT). All samples were fabricated at the Binnig and Rohrer Nanotechnology Center (BRNC) at IBM Research – Zurich.

# Optomechanics with one-dimensional gallium phosphide photonic crystal cavities: supplementary material

## 1. Simulation of the vacuum optomechanical coupling rate

For the finite-element simulations of the vacuum optomechanical coupling rate $g_0$, we consider both the moving boundary effect and the photoelastic effect, namely $g_0 = g_{0,\text{mb}} + g_{0,\text{pe}}$. We start with a first–order perturbation theory description [1], which yields

$$g_0 = -\frac{\omega_o}{2} \frac{\int E \frac{\partial \varepsilon}{\partial x} E^* \, dV}{\int E \cdot D \, dV} \quad \textbf{(S1)}$$

Here, $E$ is the electric field, $D$ is the electric displacement field, $\varepsilon$ the material permittivity, and $x$ is the generalized mechanical displacement. In the case of a shifting dielectric boundary, the contribution to $g_0$ becomes [1]

$$g_{0,mb} = -\frac{\omega_o}{2} \frac{\int (q \cdot \hat{n})(\Delta \varepsilon E_\parallel^2 - \Delta \varepsilon^{-1} D_\perp^2) \, dS}{\int E \cdot D \, dV}, \quad \textbf{(S2)}$$

where $\Delta \varepsilon = \varepsilon_1 - \varepsilon_2$ is the difference between the dielectric constants of GaP and air, $\Delta \varepsilon^{-1} = \varepsilon_1^{-1} - \varepsilon_2^{-1}$, and the subscripts $\parallel$ and $\perp$ indicate the field components parallel and perpendicular to the surface, respectively. The normalized displacement is $q$, and $\hat{n}$ is the unit vector normal to the surface $S$ between GaP and air.

The photoelastic effect in an anisotropic medium with refractive index $n$ can be described as $\frac{\partial \varepsilon}{\partial x} = -\varepsilon_0 n^4 p_{ijkl} S_{kl}$, where $\varepsilon_0$ is the vacuum permittivity. The photoelastic contribution to the optomechanical vacuum coupling rate is then

$$g_{0,pe} = -\frac{\omega_o}{2} \frac{\int E (\varepsilon_0 n^4 p_{ijkl} S_{kl}) E^* \, dV}{\int E \cdot D \, dV}, \quad \textbf{(S3)}$$

with $p$ being the photoelastic tensor and $S$ the strain tensor [2]. We use the following photoelastic coefficients: $p_{11} = -0.23$, $p_{12} = -0.13$ and $p_{44} = -0.10$ [3]. These coefficients were measured for a vacuum wavelength of $\lambda = 632.8$ nm. We note that there are further publications treating the photoelastic coefficients of GaP [4], but none presenting results at our design wavelength of $\lambda_{\text{vac}} = 1550$ nm. Because the dispersion of the photoelastic effect for wavelengths far from the bandgap is expected to be small, the chosen coefficient values should provide a reasonable estimate of the magnitude of the photoelastic contribution to the optomechanical coupling. Nevertheless, to evaluate the validity of these numbers, we additionally consider published piezooptic coefficients [5], as these can be combined with elastic stiffness coefficients to calculate the photoelastic coefficients [3]. We extrapolate the model given in [5], which was used to analyze piezooptic data over the range $\lambda_{\text{vac}} = 460$ to 1240 nm, and find that the predicted values $p_{12} - p_{11} = 0.10$ and $p_{44} = -0.08$ for $\lambda_{\text{vac}} = 1550$ nm are consistent with the photoelastic coefficients in [3].

## 2. Simulation of design tolerances

To estimate the required fabrication accuracy, we perform sweeps in the COMSOL Multiphysics simulation of the waveguide thickness, the ellipse radii and the waveguide width. The evaluated device properties are the optical and the mechanical resonance frequency, the optical quality factor, the optomechanical coupling rate and the fitness $F = g_0 \cdot Q_o$ (with $Q_o$ capped at $1.5 \times 10^6$ to avoid weighting the optimization with unattainably high values). The results shown in Fig. S1 indicate that the design is quite robust with respect to fabrication inaccuracies. The mechanical and optical resonance frequencies as well as the optical quality factor are affected most by the ellipse radii. Increasing the radii by 10 % (15 nm) leads to a drop of the quality factor by 70 %, but the simulated quality factor still exceeds one million. The optomechanical coupling rate $g_0$ exhibits a more complex dependence on the swept parameters; for the range of simulated values the variation is around $\pm 2\pi \times 100$ kHz.

## 3. Fabrication details

Device fabrication follows our previously published method for integrated GaP devices [6] with some improvements and additions. The process begins with preparation of a GaP-on-insulator substrate. An epitaxial GaP/Al$_{0.36}$Ga$_{0.64}$P/GaP heterostructure is grown by metal-organic chemical vapor deposition (MOCVD) on a single-side-polished, 2-inch, [100]-oriented GaP wafer using trimethylgallium, trimethylaluminum and tertiarybutylphosphine as precursors and a susceptor temperature of 650 °C. The initial 100-nm-thick homoepitaxial layer of GaP is deposited to improve the crystal surface quality. The subsequent Al$_{0.36}$Ga$_{0.64}$P etch-stop layer, also 100-nm thick, is required for the eventual separation of the 300-nm-thick top GaP device layer.

Following MOCVD growth, a thin (~5 nm) film of Al$_2$O$_3$ is deposited by atomic layer deposition (ALD) on both the GaP device layer and a 4-inch silicon target wafer capped with 2 µm of SiO$_2$ prepared by thermal dry oxidation at 1050 °C. The two wafers are wet cleaned to obtain hydrophilic surfaces. The GaP wafer is bonded face down onto the oxidized silicon wafer by bringing the cleaned wafers into contact at room temperature followed by annealing at 300 °C in vacuum. Approximately 350 µm of the original 400-µm thick GaP substrate wafer is removed by lapping. The bonded wafer is then diced into 7-mm × 7-mm chips. The individual chips are thinned with a fast inductively-coupled-plasma reactive ion etching (ICP-RIE) process utilizing a mixture of SiCl$_4$ and SF$_6$ [7]. The ICP-RIE process exhibits high selectivity (exceeding 1000:1), stopping abruptly on Al$_{0.36}$Ga$_{0.64}$P while maintaining an etch rate of 3 µm/min. Submersion in concentrated HCl for 90 s selectively removes the Al$_{0.36}$Ga$_{0.64}$P stop layer, yielding the desired GaP-on-SiO$_2$ substrate.

All structures (photonic crystal cavities (PhCCs) as well as waveguides and grating couplers) are patterned together in a single electron-beam lithography step using 6% hydrogen silsesquioxane (HSQ) as a negative resist (90 nm nominal thickness). Prior to spin coating with HSQ, the surface of the GaP is coated by ALD with 3 nm of SiO$_2$ for enhanced adhesion. Pattern transfer into the GaP device layer is carried out with an ICP-RIE process employing a mixture of Cl$_2$, BCl$_3$, H$_2$, and CH$_4$ [6]. After etching, the chip is treated with an oxygen plasma, and then the resist is stripped with standard 7:1 buffered oxide etchant (buffered HF).

To create freestanding PhCC devices, the patterned chips are coated with the positive resist AZ6612. The resist is exposed and developed to make rectangular openings around the optical cavities while leaving the input/output waveguides and grating couplers protected. Good adhesion of the photoresist is essential and is achieved via deposition of 5 nm of SiO$_2$ by ALD followed by treatment with hexamethyldisilazane before spin coating of the resist. Approximately 1.1 µm of the sacrificial buried SiO$_2$ layer under the devices is then removed by submerging the chip in buffered HF for 15 min.

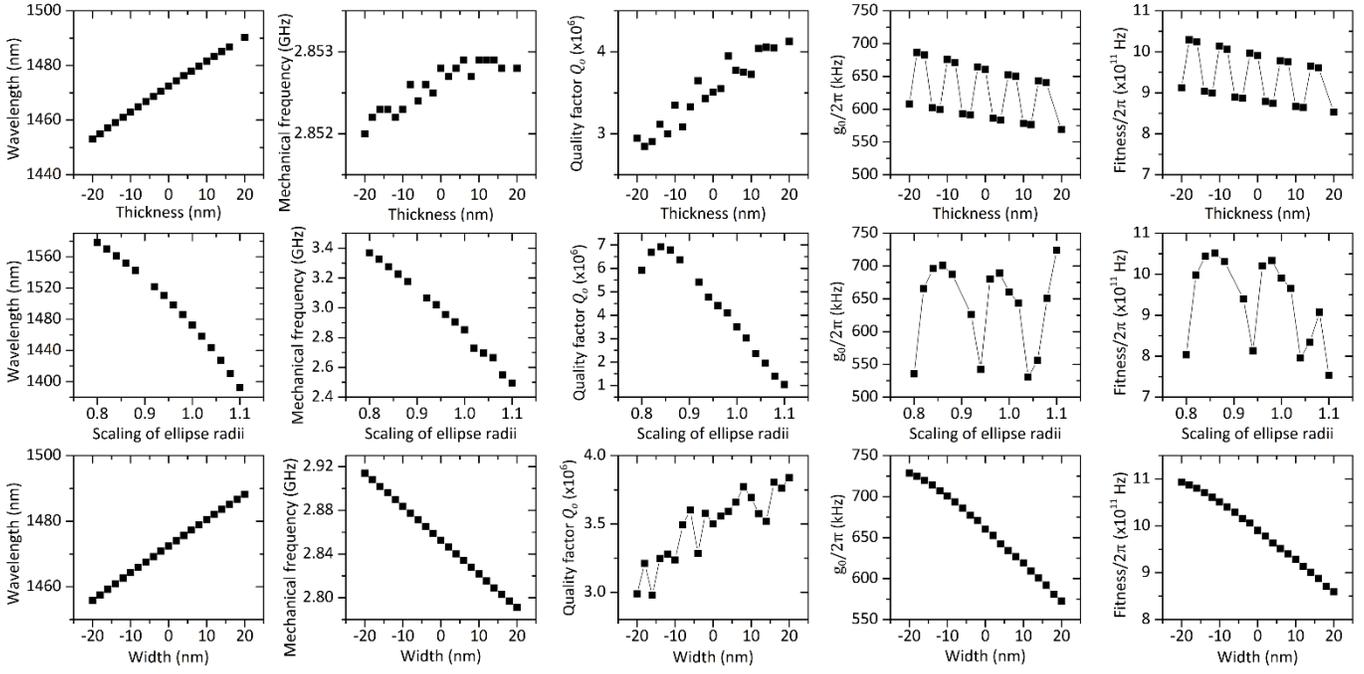

Fig. S1. Results of parameter sweeps performed in finite-element simulation of the optomechanical properties, where the waveguide thickness, the hole radii and the waveguide width were varied. Scaling of the ellipse radii is relative to the radii of the optimal design as shown in Fig. 1(a) of the main text.

## 4. Thermo-optic measurements

The optical resonance frequency of a PhCC shifts with temperature due to a combination of the thermo-refractive effect and thermal expansion. The overall frequency shift for a temperature change $\Delta T$ can be written as $\delta\omega_o = \beta(\omega_o)\Delta T$. We determine the frequency-dependent coefficient $\beta(\omega_o)$ by measuring the position of the optical resonance as a function of sample temperature at low intensity, where there is no significant thermo-optic shift due to the transmitted laser light. Specifically, the temperature of the aluminum block on which the sample chip is mounted is varied with a Peltier element and measured with an integrated thermistor. Fig. S2 shows the results for the resonance at $\lambda_{\text{vac}} = 1494$ nm of the device used for the measurements in Fig. 4 of the main text. A linear fit of the data gives a slope of $-0.00683$ THz/K. For comparison to Si, we consider the two devices characterized in reference [8], one with eight holes on each side and one with nine holes on each side, for which the values of $\beta(\omega_o)$ are $0.00756$ THz/K and $-0.00702$ THz/K, respectively.

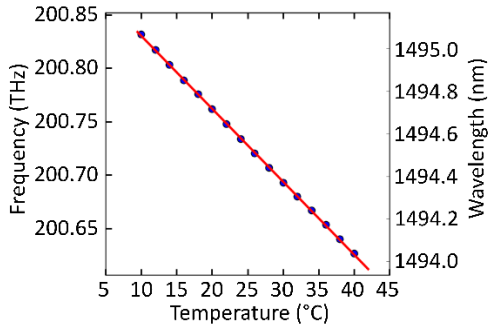

Fig. S2. Optical resonance frequency as a function of sample temperature. A linear fit (red line) gives a slope of $-0.00683$ THz/K.

We estimate the thermal resistance $R_{\text{th}}$ of a PhCC by simulating steady state heat transport with the finite element software COMSOL Multiphysics. Fig. S3 displays the geometry employed in the simulations. We take advantage of symmetry and simulate only half the structure. Heat flux enters the device through the cross-section surface at the center of the PhCC, and the boundaries indicated in Fig. S3 are set at 298 K. The thermal properties used for the respective materials are listed in Table S1. Note that we assume the devices are surround by air at constant pressure but adjust the thermal conductivity, density and heat capacity of the air as a function of local temperature. We do not take into account convection or radiation. For GaP, the PhCC has 11 holes on each side, and the dimensions are those of the device used for the measurements in Fig. 4 of the main text. In the case of Si, one device has eight holes and the other has nine holes on each side, and the dimensions are taken from [8]. The total freestanding length of the PhCC beams is 36 μm and 30 μm for the GaP and Si devices, respectively, corresponding to the values for the fabricated devices. For our GaP PhCC, we calculate a thermal resistance of $R_{\text{th}} = 1.93 \times 10^3$ K/mW. For the eight-hole and nine-hole Si devices, we estimate $R_{\text{th}} = 1.94 \times 10^3$ K/mW and $R_{\text{th}} = 1.95 \times 10^3$ K/mW, respectively. We note that the presence of holes in the PhCC has an influence on the thermal conductivity, but given the closely related designs, the similar beam lengths and the fact that the lower thermal conductivity of GaP compared to Si is compensated by a larger beam cross-section, it is not surprising that $R_{\text{th}}$ for the GaP and Si devices are nearly the same.

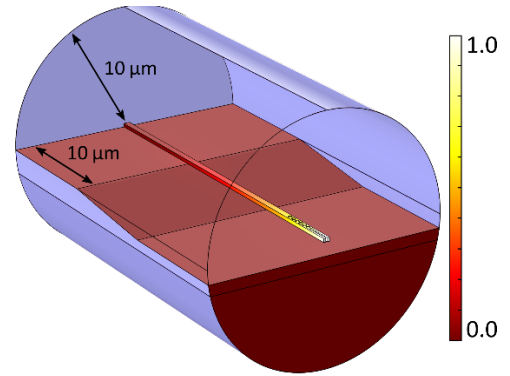

Fig. S3. Model geometry used for estimation of the thermal resistance $R_{\text{th}}$ of the GaP and Si PhCCs. The surface in blue is set at 298K (part of which has been removed to make the PhCC visible).

Table S1. Material properties used for the calculation of $R_{\text{th}}$.

| | Thermal conductivity (W/(m K)) | Density (kg/m³) | Heat capacity at constant pressure (J/(kg K)) |
|---|---|---|---|
| Si | 130 | 2329 | 700 |
| GaP | 110 | 4138 | 430 |
| SiO$_2$ | 1.38 | 2203 | 703 |

## 5. Calibration tone measurement

To determine the vacuum optomechanical coupling rate $g_0$, the thermomechanical radio frequency spectrum is compared to the signal from a calibration tone created by modulating the light entering the PhCC with a phase modulator [9]. The power spectral density of the mechanical mode is given by Eq. (S4) and that of the calibration tone by Eq. (S5).

$$S_\omega(\Omega) \approx 8 g_0^2 n_{\text{th}} \cdot \frac{\Omega_m^2}{(\Omega^2 - \Omega_m^2)^2 + \Gamma_m^2 \Omega_m^2} \quad \text{(S4)}$$

$$S_\omega^{\text{cal}}(\Omega) = \tfrac{1}{2}\Omega_{\text{cal}}^2 \beta^2 \delta(\Omega - \Omega_{\text{cal}}) \quad \text{(S5)}$$

Here, $\Omega_m$ and $\Omega_{\text{cal}}$ are the frequencies of the mechanical resonance and the calibration tone, respectively, $\Gamma_m$ denotes the mechanical linewidth, $n_{\text{th}} \simeq k_B T / \hbar \Omega_m$ is the number of phonons occupying the mechanical mode at the ambient temperature $T$, with $k_B$ as Boltzmann's constant, and $\beta$ is the modulation depth applied to create the calibration tone. The power spectral density from the photoreceiver for both signals is given by $S_V(\Omega) = |G_{V\omega}(\Omega)|^2 \cdot S_\omega(\Omega)$, where $G_{V\omega}(\Omega)$ is a frequency dependent transduction factor. Comparing the areas beneath the calibration tone, $\langle V^2 \rangle_{\text{cal}} = \tfrac{1}{2}\Omega_{\text{cal}}^2 \beta^2 |G_{V\omega}(\Omega_{\text{cal}})|^2$, and the thermomechanical noise peak, $\langle V^2 \rangle_m = 2 g_0^2 n_{\text{th}} |G_{V\omega}(\Omega_{\text{cal}})|^2$, we can determine $g_0$ from Eq. (S6).

$$g_0 = \frac{\beta \Omega_{\text{cal}}}{2} \sqrt{\frac{1}{n_{\text{th}}} \frac{\langle V^2 \rangle_m}{\langle V^2 \rangle_{\text{cal}}}} \left| \frac{G_{V\omega}(\Omega_{\text{cal}})}{G_{V\omega}(\Omega_m)} \right| \quad \text{(S6)}$$

Choosing $\Omega_m$ and $\Omega_{\text{cal}}$ to differ by only a few megahertz, we can assume the transduction factor to be nearly the same, and therefore $\left| \frac{G_{V\omega}(\Omega_{\text{cal}})}{G_{V\omega}(\Omega_m)} \right| \approx 1$.

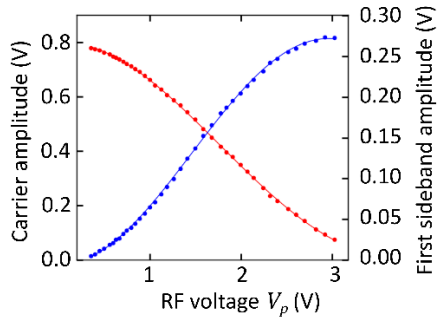

Fig. S4. Measured amplitude of the carrier (red dots) and first-order sidebands (blue dots) created by modulation of the laser light at 1550 nm with a phase modulator driven at 2.9 GHz and amplitude $V_p$. The curves are fits to the data and give a half-wave voltage $V_\pi = 5.09 \pm 0.02$ V.

To ascertain the modulation depth $\beta = V_p / V_\pi \cdot \pi$ of the calibration tone, where $V_p$ is the amplitude of the radio frequency voltage applied to the phase modulator, the half-wave voltage $V_\pi$ of the modulator must be known. We measure $V_\pi$ by setting the radio frequency drive signal to 2.9 GHz and varying $V_p$. The light transmitted through the phase modulator is guided to a scanning Fabry-Pérot interferometer (Thorlabs SA210-12B), which resolves the carrier and the sideband signals. The carrier amplitude as a function of $V_p$ can be described as $A_0 = a \left( J_0(\pi V_p / V_\pi) \right)^2$ and the first sideband amplitude as $A_1 = a \left( J_1(\pi V_p / V_\pi) \right)^2$, where $J_0$ and $J_1$ are Bessel functions of the first kind and $a$ is a scaling factor. A fit of these functions to the measured carrier and sideband amplitudes (see Fig. S4) yields a half-wave voltage $V_\pi = 5.09 \pm 0.02$ V, in agreement with the vendor specifications (4.5 V $< V_\pi <$ 6.5 V for 3 GHz).

Fig. S5 shows an example of a spectrum recorded with the calibration-tone method for the mechanical mode near 2.9 GHz. For this device, an optomechanical vacuum coupling rate of $g_0 = 2\pi \times (370 \pm 40)$ kHz is inferred.

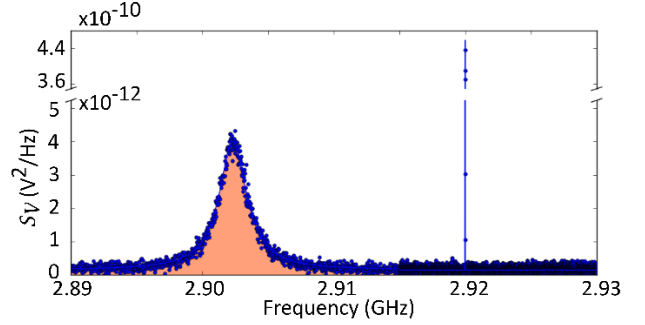

Fig. S5. Power spectral density of the mechanical resonance (left peak) and the calibration tone (right peak). An optomechanical vacuum coupling rate of $g_0 = 2\pi \times 370 \pm 40$ kHz is inferred.

## 6. Optomechanically induced absorption

To observe optomechanically induced absorption (OMIA), the laser control field at frequency $\omega_c$ is blue detuned from the cavity resonance at frequency $\omega_o$. Additional probe fields much weaker than the control field are created by phase modulating the laser. In a frame rotating at $\omega_o$, where $\Delta_{oc} = \omega_o - \omega_c$ is the laser-cavity detuning and $\Delta_{pc} = \omega_p - \omega_c$ is the probe-control detuning for a probe frequency $\omega_p$, the expected transmission is [10]

$$t_p(\Delta_{pc}) = \frac{\kappa_e / 2}{i(\Delta_{oc} + \Delta_{pc}) + \kappa/2 + \frac{G^2}{i(\Omega_m - \Delta_{pc}) - \Gamma_m/2}}. \quad \text{(S7)}$$

Here, $\kappa$ is the total optical decay rate, $\kappa_e$ is the contribution to the decay rate from loss through the Bragg mirrors of the PhCC, and $G = g_0 \sqrt{n_{\text{cav}}}$ is the pump-enhanced optomechanical coupling rate for $n_{\text{cav}}$ intracavity photons. The electric field transmitted through the PhCC contains components oscillating at the carrier and both sideband frequencies:

$$E_{\text{out}} = e^{i\omega_c t} \left\{ t_p(0) + t_p(\Delta_{pc}) \tfrac{\beta}{2} e^{i\Delta_{pc} t} + t_p(-\Delta_{pc}) \tfrac{\beta}{2} e^{-i\Delta_{pc} t} \right\} \quad \text{(S8)}$$

where $t$ is time, and $\beta$ is the modulation depth for the probe sidebands. The signal detected at the photodiode is proportional to $|E_{\text{out}}|^2$. The component oscillating with frequency $\Delta_{pc}$ is

$$I_{\Delta_{pc}} \propto \left\{ |t_p(-\Delta_{pc})| \cos(\Delta_{pc} t + \varphi_-) + |t_p(\Delta_{pc})| \cos(\Delta_{pc} t + \varphi_+) \right\}. \quad \text{(S9)}$$

The phase shifts experienced by the lower and upper sidebands are $\varphi_-$ and $\varphi_+$, respectively. Ideally $\varphi_+ - \varphi_- = \pi$, but in practice the two sidebands acquire an additional relative phase shift [11], so $\varphi_-$ is replaced by $\varphi_+ + \theta$, and $\varphi_+$ and $\theta$ are used as fit parameters. The vector network analyzer (VNA) measures the in-phase (Eq. (S10)) and quadrature (Eq. (S11)) components of the signal

$$I = |t_p(-\Delta_{pc})| \cos(\varphi_-) + |t_p(\Delta_{pc})| \cos(\varphi_+) \quad \text{(S10)}$$

$$Q = |t_p(-\Delta_{pc})| \sin(\varphi_-) - |t_p(\Delta_{pc})| \sin(\varphi_+) \quad \text{(S11)}$$

The $|S_{21}|$ parameter is proportional to $\sqrt{I^2 + Q^2}$.

Fitting the above model to the $|S_{21}|$ response of the VNA, we can extract various values, including the photon enhanced coupling rate $G$. To determine the vacuum coupling rate $g_0 = G/\sqrt{n_{\text{cav}}}$, the intracavity photon number $n_{\text{cav}}$ must be known. It can be estimated from the power leaving the cavity using $P_{\text{out}} = n_{\text{cav}} \hbar \omega_o^2 / 2 Q_e$. We calculate the external optical quality factor $Q_e$ by evaluating the transmission for a known detuning $\Delta_{oc}$ and combine the result with the measured total optical quality factor $Q_o$ [12] using Eq. (S12).

$$T(\Delta_{oc}) = \frac{P_{\text{out}}}{P_{\text{in}}} = \frac{1/4Q_e^2}{(\Delta_{oc}/\omega_o)^2 + 1/4Q_o^2} \tag{S12}$$

Because the power entering and leaving the cavity cannot be measured directly before and after the device, but only at positions separated from the cavity by various lossy components such as the grating couplers, $Q_e$ and therefore $n_{\text{cav}}$ can only be determined with limited accuracy. We estimate the grating-coupler losses by measuring structures in which the PhCC has been replaced with a simple straight waveguide.

## 7. Determination of laser detuning for evaluation of dynamical backaction

Whenever laser light passes through the PhCC, the thermo-optic effect causes a shift of the optical resonance, which depends on the intracavity energy, as shown in Fig. S6(a). As the laser frequency is varied, the optical resonance frequency will change, and the detuning of the laser with respect to the cavity $\Delta_{\text{oc}}$ is not obvious. To determine the detuning, as is required for evaluation of the optical spring effect and parametric amplification, we make use of two independent methods.

In the first, the transmitted optical power is recorded for each thermomechanical radio frequency spectrum taken. Assuming a Lorentzian transmission profile and using the known values for optical linewidth and maximum transmitted optical power, we assign a detuning to each transmitted power value. The resulting dependence of detuning on transmitted power is shown in Fig. S6(b) with blue open circles.

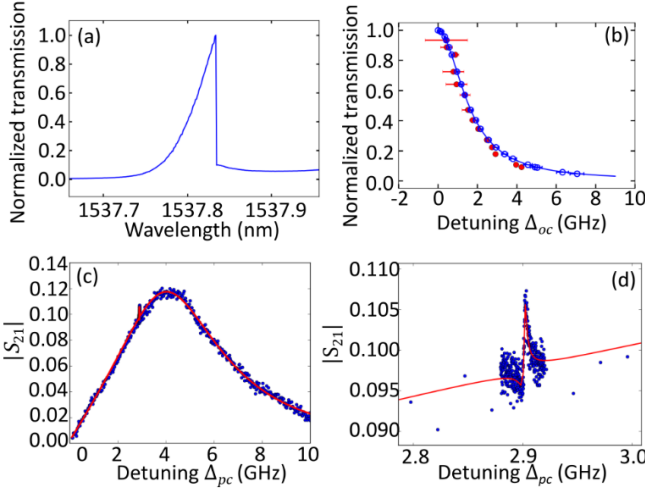

Fig. S6. (a) Transmission profile of the device used for measurement of dynamical backaction (21 μW input power). Due to the thermo-optic shift, the position of the optical resonance changes while sweeping the laser wavelength, producing a triangular transmission profile. (b) Plot of detuning inferred from transmission (blue open circles) assuming a Lorentzian profile with linewidth $\kappa = 2\pi \times 3.15$ GHz (blue line) and from fits of OMIA traces (red closed circles and error bars). (c) and (d) Example of an OMIA trace (blue dots) with fit to Eq. (S7) (red line), where (d) is a magnification of the area around the OMIA peak in (c).

For the second method, we recorded a VNA trace as described for the measurements of OMIA, the shape of which depends trace strongly on the detuning (see Fig. S6(c) and (d) for an example). Fitting Eq. (S7) to the trace gives the detuning $\Delta_{\text{oc}}$. This method does not work for very large or very small detuning, as the OMIA peak is then no longer visible.

The results are plotted as red solid circles in Fig. S6(b) and overlap well with the values obtained from the first method. For the further evaluation of dynamical backaction, we used the average of the values obtained from the two methods.